\documentclass[aps,pra,showpacs,twocolumn]{revtex4}
\usepackage{graphicx}
\usepackage{amsmath}

\def\ket{\rangle}
\def\<{\langle}
\def\>{\rangle}

\begin{document}

\title{Proof of Security of a High-Capacity Quantum Key Distribution Protocol}
\author{Xiaowei Zhang$^1$, Kai Wen$^1$ and Gui Lu Long$^{1,2}$}
\address{$^1$Key Laboratory for Quantum Information and Measurements and
 Department of Physics, Tsinghua University, Beijing
100084, P. R. China\\
$^2$ Key Laboratory of Atomics and Molecular Nanosciences,
Tsinghua University, Beijing 100084, P R China}
\date{\today}

\begin{abstract}
 We prove the security of a high-capacity quantum key distribution
protocol over noisy channels. By using entanglement purification
protocol, we construct a modified version of the protocol in which
we separate it into two consecutive stages. We prove their
securities respectively and hence the security of the whole
protocol.
\end{abstract}

\maketitle

\section{Introduction}
\label{s-intro}
 Security has been one of the most important concerns ever since
people began to communicate. Many classical cryptographic
protocols are based on the computational infeasibility of some
mathematical problems such as the factorization of large composite
numbers, which will be solved in a quantum computer using Shor's
algorithm\cite{Shor94}. Quantum cryptography, on the other hand,
relies on the principles of quantum mechanics, especially the
uncertainty principle and the no-cloning theorem\cite{no-cloning},
therefore is provably secure. The quantum cryptography has
captured more and more attention. It is mainly used in
establishing secrete key between two parties. The scheme is as
follows: two participants, commonly called Alice and Bob, share a
quantum channel. They transmit quantum states (generally refer to
as qubits) which encode classical information, and measure the
qubits in certain bases to get key code. Since interference
introduced by the eavesdropper, commonly known as Eve, will
disturb quantum states and can be detected, Alice and Bob will
share their messages while leaking little information to Eve. Many
quantum key distribution protocols (QKD) have been proposed since
the first protocol\cite{BB84} published by Bennet and Brassard,
for instance in protocols in Refs. \cite{EPR1, EPR2, B92, gv95,
six-state, high-capacity, long0, longa1}.

 Security and efficiency are two principal factors in QKD
protocols. Since both the channel noise and eavesdropping can
disturb the quantum states, a good definition of security requires
a distinct differentiation between them. One parameter of security
serves this purpose. Denoted as the tolerable bit error rate, it
is defined as follows: below this threshold, a QKD is secure by
using quantum error correction and privacy amplification. The fast
development of QKD protocols requires explicit proof of their
securities over noisy channels. Mayers's\cite{Mayers} and
Biham's\cite{Biham} work reach this end by complex calculation. A
different approach, by using entanglement purification protocols
(EPP) which can purify the EPR pairs by sacrificing some of
them\cite{bdsw}, was proposed\cite{Lo-Chau-science}. After that,
many proofs with this basic idea have been given\cite{six-state,
Shor-Preskill, twoway, practical_twoway}.

 Efficiency is also an important parameter of QKD protocols. In this
paper, we will concentrate on a theoretically high-capacity QKD
protocol\cite{high-capacity}. Its efficiency is achieved by
adopting EPR pair which encodes 2 classical bits.

 The high-capacity protocol transmits an EPR sequence in two steps, one
particle sequence at a time, thus is protected against
eavesdropping in ideal quantum channel (that is, channel without
noise). In order to make it error-resistant in noisy quantum
channels, we apply the method of Shor and
Preskill\cite{Shor-Preskill} and add an EPP to the protocol to
construct a modified version so as to prove its security. Our
paper is organized as follows: in section \ref{s-nota}, we specify
the notations used in this paper. In section \ref{s-review}, we
briefly review the high-capacity protocol and outline the proof of
security in ideal situations. In section \ref{s-first}, we add
security check procedures to the original protocol and prove the
first stage of the modified version is secure. In section
\ref{s-second}, we prove the security of the second stage of the
modified high-capacity protocol. Then we prove that this modified
version is equivalent to the original one and give a brief summary
in section \ref{s-conclusion}.

\section{Notation}
\label{s-nota}
 The notations used here are mostly the same as that of the
high-capacity protocol\cite{high-capacity} and that of the Shor and
Preskill's\cite{Shor-Preskill}.

 Bell bases are the four maximally entangled states:
\begin{equation}
|\Phi^\pm\ket = \frac{1}{\sqrt{2}} (|00\ket\pm|11\ket),
|\Psi^\pm\ket = \frac{1}{\sqrt{2}} (|01\ket\pm|10\ket).
\end{equation}
We use $|\Phi^{+}\ket$, $|\Phi^{-}\ket$, $|\Psi^{+}\ket$ and
$|\Psi^{-}\ket$ to represent 00, 01, 10 and 11 respectively.

 The three Pauli matrices are:
\begin{equation}
\sigma_{x}= \left(\begin{array}{cc}\ 0 & \ 1 \\ \ 1 & \
0\end{array}\right), \sigma_{y}= \left(\begin{array}{cc}\ 0 & \ -i \\
\ i & \ 0\end{array}\right), \sigma_{z}= \left(\begin{array}{cc}\ 1
& \ 0 \\ \ 0 & \ -1\end{array}\right).
\end{equation}
They are used in the error checking process.

 The Hadamard transform, $H$, is of the form:
\begin{equation}
H=\frac{1}{\sqrt{2}} \left(\begin{array}{cc}\ 1 & \ 1 \\ \ 1 & \
-1\end{array}\right).
\end{equation}
It interchanges the basis $|0\ket$, $|1\ket$ and $|+\ket$, $|-\ket$,
where $|+ \ket = \frac{1}{\sqrt2}(|0\ket+|1\ket)$ and $|- \ket =
\frac{1}{\sqrt{2}}(|0\ket-|1\ket)$. In what we will see in section
\ref{s-first}, after randomly performs the Hadamard gate, bit flip
errors and phase flip errors would be uncorrelated.

 We introduce the Calderbank-Shor-Steane(CSS) code in the entanglement
purification process, a CSS code\cite{Shor-Preskill} is defined as
follows: $C_1$ and $C_2$ are two classical binary codes which
satisfy
\begin{equation}
\{0\} \subset C_{2} \subset C_{1} \subset F^{n}_{2}
\end{equation}
where $F^{n}_{2}$ is the binary space of $n$ bits. $C_{1}$ and
$C^{\bot}_{2}$ can correct up to $t$ bit errors. A basis of the CSS
code can be built as follows, for $v \in C_{1}$, define the vector
\begin{equation}
v \to \frac{1}{|C_{2}|^{1/2}} \sum_{\omega \in C_{2}} |v+ \omega
\ket
\end{equation}
Notice that when $v_{1} - v_{2} \subset C_{2}$, they give the same
code. So the CSS code corresponds to the coset of $C_{2}$ in
$C_{1}$. Let $H_1$ and $H_2$ be the parity check matrix for the code
$C_1$ and $C^{\bot}_2$ respectively. Then by measuring
$\sigma^{[r]}_{z}$ for each row $r\in H_1$ and $\sigma^{[r]}_{x}$
for each row $r\in H_2$, the CSS code can correct up to $t$ bit
errors.

\section{Review of the high-capacity protocol}
\label{s-review}
 The key point of the high-capacity protocol is that it
uses each EPR pair to encode 2 bits of key code. By building an
ordered EPR sequence and sending each half in two steps from Alice
to Bob, Alice and Bob can protect their communication against most
eavesdropping attacks.

Here we outline the high-capacity protocol\cite{high-capacity}.

\emph{Protocol 1: Theoretical efficient high-capacity Protocol}
\begin{enumerate}
\item Alice produces an ordered $N$ EPR pair sequence: $ [(P_1(1)$,
$P_1(2))$, $(P_2(1)$, $P_2(2))$, $\ldots$, $(P_i(1)$, $P_i(2))$,
$\ldots, $ $(P_N(1),$ $P_N(2))]$.

\item Then Alice takes one particle from each EPR pair to form two
ordered EPR partner particle sequences: $[P_1(1)$, $P_2(1)$,
$\ldots$, $P_N(1)]$ and $[P_1(2)$, $P_2(2)$, $\ldots$, $P_N(2)]$.
Alice sends to Bob one ordered EPR partner particle sequence:
$[P_1(2)$, $P_2(2)$, $\ldots$, $P_N(2)]$.

\item After Bob receiving the ordered EPR partner particle
sequence, randomly he chooses a sufficiently large subset of his
sequence and performs measurement on the particles in the subset
randomly in of the two measuring-basis $\sigma_z$ or $\sigma_x$.
The result of this measurement will be either 0 or 1. Bob stores
the rest of the particles of his EPR particle sequence.

\item Then Bob tells Alice through a classical channel his reception
of the particle sequence and the particles that he has chosen to
measure in a certain direction.

\item (First eavesdropping check.)
After hearing from Bob, Alice then performs measurement on the
partner subset of those particles whose partner has been measured by
Bob. They then publicly compare their results of these measurements
to check eavesdropping.

\item If they are certain that there is no eavesdropping, then Alice
sends Bob the remaining EPR particle sequence: $[P_1(1)$, $P_2(1)$,
$\ldots$, $P_N(1)]$.

\item After Bob receives these $N$ particles, he first drops
the particles that have been measured, then takes one particle from
each particle sequence in order and performs Bell-basis measurement
on them. He records the results of the measurements.

\item (Second eavesdropping check.)
Alice and Bob choose a sufficiently large subset of these Bell-basis
measurement results to determine if the QKD is successful. If the
error rate in this check is below a certain threshold, then the
results are taken as raw key.
\end{enumerate}

 Unlike many other QKD protocols\cite{BB84, EPR1, EPR2}, in which an
EPR pair is used to encode one key bit, the high-capacity protocol
here encodes two bits, thus the capacity is doubled.

 As discussed in the original paper\cite{high-capacity}, this protocol
is secure in quantum channels without noise.

\section{The first stage of the modified high-capacity protocol}
\label{s-first}
 From the review of the high-capacity protocol, we see that it
is an efficient protocol, but a critical questions remains: how can
this protocol protect against quantum channel noise? In other words,
how can Bob distinguish between channel noise and eavesdropping, and
can correct bit errors when necessary?

 In this section, we add entanglement purification steps to the
first stage of the high-capacity protocol and prove its security
over noisy channels.

\emph{Protocol 2: Secure high-capacity Protocol(stage 1)}
\begin{enumerate}
\item Alice produces an ordered $N$ EPR pair sequence according to
a quaternary string $a$. More specifically, she creates
$|\Phi^{+}\ket$, $|\Phi^{-}\ket$, $|\Psi^{+}\ket$ or $|\Psi^{-}\ket$
when the corresponding $a_i=0, 1, 2$ or $3$. Thus build $ [(P_1(1)$,
$P_1(2))$, $(P_2(1)$, $P_2(2))$, $\ldots$, $(P_i(1)$, $P_i(2))$,
$\ldots$, $(P_N(1),$ $P_N(2))]$.

\item Then Alice takes one particle from each EPR pair to form two
ordered EPR partner particle sequences: $[P_1(1)$, $P_2(1)$,
$\ldots$, $P_N(1)]$ and $[P_1(2)$, $P_2(2)$, $\ldots$, $P_N(2)]$.
She retains the first sequence.

\item Alice randomly chooses the bases of the second half of the
EPR pairs. More specifically, Alice selects a random $N$ binary
string $b$ and applies Hadamard transformation on $P_i(2)$ when the
corresponding $i$th bit is 1.

\item Alice sends the second EPR sequence to Bob.

\item Bob receives the $N$ qubits and publicly announces the
reception.

\item Alice randomly chooses $n(n<N)$ bits as checking bits and
leaves the rest $N-n$ bits unchanged.

\item Alice then tells Bob the bit string $b$ and which $n$ bits
are checking bits.

\item Bob performs Hadamard transformation on the qubits where the
corresponding components of $a$ are 1. They measure the $n$ checking
qubits in $|0\ket, |1\ket$ bases, if too many outcomes disagree,
they abort the protocol.

\item Alice and Bob applies a CSS code  and makes
$\sigma^{[r]}_{z}$ according to each row $r\in H_1$ and
$\sigma^{[r']}_{x}$ according to each row $r'\in H_2$ to their EPR
particles. They compute the syndromes and make corrections in
order that they obtain $k(k<N-n)$ nearly perfect EPR pairs.
\end{enumerate}

 The above protocol is the first stage of the modified version of the
high-capacity protocol. We employ CSS code here to purify the
entangled states. Notice that the vector space $C_1$ and
$C^{\bot}_2$ are orthogonal, so $\sigma^{[r]}_{z}$ and
$\sigma^{[r']}_{x}$ commutes. The measurement computes the error
syndrome for bit flip and phase flip respectively, then after the
measurement, Alice and Bob can obtain $k$ perfect EPR pairs.

 We next show that the bit and phase errors are uncorrelated. As noted
by Lo and Chau\cite{six-state}, for a quantum channel, the error
rate can be expressed by a density matrix $diag(a, b, c, d)$, in
which $a$, $b$, $c$ and $d$ represent the probabilities of zero
errors, bit flip errors, phase flip errors and both bit and phase
flip errors. As noted in section \ref{s-nota}, the Hadamard
transformation interchanges the basis $|0\ket$, $|1\ket$ and
$|+\ket$, $|-\ket$, so it changes bit flip error to phase flip
error and vice versa. When $N$ is large enough, there is a high
probability to have $\frac{N}{2}$ 1 in the binary string $b$, so
there are almost half Hadamard transforms operating on the EPR
pairs. Averaging over the two cases of Identity and Hadamard, the
effective density matrix shared by Alice and Bob after the
operation is $diag(a, \frac{b+c}{2}, \frac{b+c}{2}, d)$. From this
matrix, we can see that the Hadamard transformation makes the bit
flip error and the phase flip error with equal probability and
uncorrelated.

 Because this modified high-capacity protocol uses CSS code, it can
successfully correct quantum state which differs from the input
state (we denoted as $|\psi\ket$) less than $t$ bit flip errors
and $t$ phase flip errors. Since all the measurements in this
protocol commute under the Bell states, we can use the classical
method to calculate the fidelity of the purified $k$ EPR pairs. In
the transmission, the error rates for the check bits and code bits
are almost the same. And because Eve doesn't know anything about
the check bits and the code bits, her interference is also the
same to these two sets. Then as $(N-n) \to \infty$, we can use
classical probability theory to calculate the fidelity, which will
yield $F(\rho, \psi)\geq 1-2^{-s}$.

 Applying the Lemma 1 and 2 of Lo and Chau\cite{Lo-Chau-science}, we
know that if Alice and Bob share a state with fidelity greater than
$1-2^{-s}$ with input state $\rho$, then Eve's mutual information
with the key would be exponentially small, so the first stage of the
modified version of the high-capacity protocol is safe.

\section{The second stage of the modified high-capacity protocol}
\label{s-second}
 In this section, we prove that the second stage of the protocol is
secure too. This stage can be seen as a repetition of the first
stage, with a some variation as we list below.

\emph{Protocol 2: Secure high-capacity Protocol(stage 2)}
\begin{enumerate}
\setcounter{enumi}{9}
\item Alice has one ordered half of $k$ perfect EPR particle
sequence: $[P'_1(1)$, $P'_2(1)$, $\ldots$, $P'_k(1)]$ and Bob has
the corresponding sequence: $[P'_1(2)$, $P'_2(2)$, $\ldots$,
$P'_k(2)]$.

\item Alice randomly chooses the bases of her own halves of the
EPR pairs. To be more specific, Alice selects a random $k$ binary
string $b'$ and applies Hadamard gates on $P_i(1)$ when the
corresponding $i$th bit is 1.

\item Alice sends her EPR sequence to Bob.

\item After receiving the sequence, Bob publicly announces this
fact.

\item Alice then tells Bob the binary string $b'$.

\item Bob selects a sufficiently large subset among his EPR pairs
and measures them according to $b'$, if too many results
inconsistent, he aborts the communication.

\item Bob performs entanglement purification in the same way as
in stage 1 to the remaining EPR pairs to obtain $m$ perfect EPR
pairs.

\item Bob performs Bell-basis measurement to obtain the secret key.
\end{enumerate}

 The proof of security of the stage 2 of protocol 2 is similar to that of
stage 1. So as discussed above, Eve can know nothing about the pure
EPR pairs in Bob's hand, then it is obvious that stage 2 is also
unconditionally secure.

 Combining the previous two stages together, we will see that by
adding entanglement purification protocol to the high-capacity
protocol will protect it against quantum channel noise.

\section{Conclusion and further discussion}
\label{s-conclusion}

We have given the security proof of the high-capacity QKD protocol
by using the Shor-Preskill method. The proof is divided into two
stages. When the error rate is below the threshold value, 11\% for
the one-way communication protocol as is true in our case, Alice
and Bob can still obtain a secure key by using entanglement
purification.

This work is supported by the National Fundamental Research
Program Grant No. 001CB309308, China National Natural Science
Foundation Grant No. 10325521, 60433050, and the SRFDP program of
Education Ministry of China.

\end{document}